\documentstyle[preprint,aps]{revtex}

\begin{document}

\title{RPA-CPA theory for magnetism in disordered Heisenberg binary systems with long range
exchange integrals.}

\author{G.~Bouzerar and P.~Bruno
\\
Max-Planck-Institute f\"ur Mikrostrukturphysik, Weinberg 2,
D-06120 Halle, Germany}
\maketitle
\begin{abstract}
We present a theory based on Green's function formalism to study magnetism in disordered Heisenberg systems with long range exchange integrals. Disordered Green's function are decoupled within Tyablicov scheme and solved with a CPA method. The CPA method is the extension of Blackmann-Esterling-Beck approach to system with environmental disorder term which uses cumulant summation of the single-site non crossing diagrams. The crucial point is that we are able to treat simultaneously and self-consistently the RPA and CPA loops. It is shown that the summation of s-scattering contribution can always be performed analytically. While the p,d,f .. contributions are difficult to handle in the case of long-range coupling. To overcome this difficulty we propose and provide a test of a simplified treatment of these terms. In the case of 3D disordered nearest-neighbor Heisenberg system, a good  agreement between the simplified treatment and the full calculation is achieved. Our theory allows in particular to calculate the Curie temperature, the spectral functions and the temperature dependence of the magnetization of each constituant as a function of concentration of impurity. Additionally it is shown that a virtual crystal treatment fails even at low impurity concentration.
\end{abstract}

PACS numbers: 75.10.-b, 75.25.+z, 71.10.-w, 75.50.Cc

\section{Introduction.}
The coherent potential approximation (CPA) is widely used to study the effect of disorder in crystals (for reviews see \cite{Elliott,Morigaki}).
The CPA was initially developed independently by Soven \cite{Soven} and Taylor
\cite{Taylor} to study systems with only {\it diagonal} disorder. 
Using a $2 \times 2$ 
formulation, a generalization to the presence of {\it off-diagonal} disorder was provided by Blackmann Esterling and Berck (BEB) \cite{BEB,Gonis}. In these approaches the main idea is to replace the system by an effective medium which is determined by the condition that the averaged T-matrix of a single impurity immersed in the effective medium is zero.
An alternative approach is based on cumulant expansion \cite{Yonezawa,Leath}.
This latter method has the advantage that it can handle the {\it environmental} disorder term which is characteristic of the Goldstone's systems (phonons, magnons).
The first proper treatment of the environmental disorder term, by using the cumulant expansion method is due to Lage and Stinchcombe \cite{LS} who studied the diluted Ising problem (S=1/2). Later, using the 2x2 matrix method of BEB, the method was extended by Whitelaw \cite{Whitelaw} to the phonon problem. In their calculations the coupling and locator are fixed quantities and restricted to nearest neighbor exchange couplings. It is well known that magnetism in clean ferromagnetic systems can be tackled with Green's function formalism using Tyablicov decoupling procedure (RPA). This method goes beyond a simple mean field since it includes quantum fluctuations. Additionally, it fulfills the Goldstone and Mermin-Wagner theorems which is not the case of a mean field treatment.
 In the case of clean systems, combining first principle calculations to evaluate the exchange integrals and RPA method it was shown that one can provide satisfactory Curie temperature for Co and Fe \cite{Pajda}. Whilst, a simple mean field calculation largely overestimate the Curie temperature.
It is our  objective to provide in this paper a generalization of the RPA method to the disordered systems. 
We show that by combining in a self-consistent manner the RPA method and the CPA treatment of the disorder we are able to calculate Curie temperature, magnetization of the different constituants, spectral weights.... The CPA treatment is done in a similar way as done by  Lage and Stinchcombe and by Whitelaw. 
However, due to Tyablicov decoupling scheme for the disordered Green's functions, the locators and the effective exchange integrals are temperature dependent and have to be determined self-consistently for a given temperature.

The paper is organized as follows. In the first section we derive 
after Tyablicov decoupling scheme the disordered binary alloy Green's function which includes {\it diagonal}, {\it off-diagonal} and {\it environmental} disorder.
In section II, we perform the calculation of the averaged Green's functions for A (respectively B) atom. In section III, by generalizing Callen's formula we derive the equations for the magnetizations $m_{A}$, $m_{B}$ and for the Curie Temperature. In section IV, we propose an alternative simplified treatment of the p,d,..scattering contribution to the self-energy to the case of system with long-range exchange coupling. Finally in section V we present some numerical results and proceed to a test of our approximation of the self-energy contribution of the higher scattering terms.

\section{Disordered Green's function and RPA decoupling scheme.}
We study the magnetism in a binary alloy $A_{1-c}B_{c}$, A and B can be either magnetic ions or paramagnetic. We denote their spin respectively $S_{A}$ and $S_{B}$.
The total Hamiltonian reads,

\begin{equation}
\hat {H}=\sum_{ij}-J_{ij}{\bf S}_{i} \cdot {\bf S}_{j} - \sum_{i}D_{i}(S^{z}_{i})^2
-B \sum_{i} g \mu_{i}(S^{z}_{i})
\end{equation}
where the $J_{ij}$ and $D_{i}$ are random variables:$J_{ij}=J^{\lambda \lambda'}_{ij}$ with the probability $P_{i}^{\lambda} P_{j}^{\lambda'}$ where $P_{i}^{\lambda}$ is the probability that the site i is occupied by a $\lambda$-atom. Similarly $D_{i}=D_{\lambda}$ with probaility $P_{i}^{\lambda}$.
 The exchange integrals are assumed to be long range, our study is not restricted to the nearest neighbor Heisenberg model. The second term which describes anisotropy is only relevant in the case of 2D systems 
to get a non zero Curie temperature $T_{c}$ (Mermin-Wagner theorem). However in the case of 3D systems the contribution of this term can be neglected. We also include the effect of an external magnetic field.

Let us consider the following retarded Green's function,

\begin{equation}
G^{+-}_{ij}(t)=-i\theta (t) \langle [S^{+}_{i}(t), S^{-}_{j}(0)] \rangle   
\end{equation}

where $\langle ..\rangle $ denotes the statistical average at temperature T,
\begin{equation}
\langle \hat{O}\rangle=\frac{1}{Z} Tr(e^{-\beta \hat{H}} \hat{O})
\end{equation}
where $Z=Tr(e^{-\beta \hat{H}})$.

$G^{+-}_{ij}(t)$'s Fourier transform in Energy space is,
\begin{equation}
\ll S^{+}_{i};S^{-}_{j} \gg =G^{+-}_{ij}(\omega)= \int_{-\infty}^{+\infty} G^{+-}_{ij}(t)e^{i \omega t} dt
\end{equation}
Its equation of motion reads,
\begin{equation}
\omega G^{+-}_{ij}(\omega)= 2 m_{i}\delta_{ij} + \ll [S^{+}_{i},H];S^{-}_{j} \gg
\end{equation}
where $m_{i}=<S_{i}^{z}>$, or $m_{i}=m_{A}$ (resp. $m_{B}$) if $i=A$ (resp. $i=B$). 

After expanding the second term on the right side of the equality we obtain,

\begin{eqnarray}
(\omega -g\mu_{i}B) G^{+-}_{ij}(\omega)= 2m_{i}\delta_{ij}\nonumber \\ 
-\sum_{l}J_{il}\ll S^{z}_{i}S^{+}_{l}-S^{+}_{i}S^{z}_{l};S^{-}_{j} \gg\nonumber \\ 
+D_{i} \ll S^{z}_{i}S^{+}_{i}+S^{+}_{i}S^{z}_{i} \gg
\end{eqnarray}

The next step consists in decoupling the higher order Green's function.
For the second term we use the standard Tyablicov decoupling \cite{Tyablicov} (equivalent to RPA). The last term due to anisotropy is somehow more complicate since on-site correlation are involved. Following the approach discussed in Ref. \cite{Jensen} we adopt for this term the Anderson-Callen decoupling scheme \cite{Anderson-Callen}:

\begin{eqnarray}
D_{i} \ll S^{+}_{i}S^{z}_{i}+S^{z}_{i}S^{+}_{i} \gg = 2D_{i}\gamma_{i} m_{i}
\end{eqnarray}
where,
\begin{equation}
\gamma_{i}=1-\frac{1}{2S^2}(S_i(S_i+1)-<(S_{i}^{z})^2>)
\label{aniso}
\end{equation}
After simplification we find,

\begin{equation}
G^{+-}_{ij}=g_i \delta_{ij}+g_i \sum_{l} \Phi_{il}G^{+-}_{lj}
-\epsilon g_i(\sum_{l} \Psi_{il})G^{+-}_{ij}
\label{propag}
\end{equation}

where $\phi_{il}=-1/2 J_{il}$ and $\Psi_{il}=-1/2 J_{il} \frac{m_{l}}{m_{i}}$
and $g_i$ denotes the locator: $g_{i}=g^{0}_{A}$ (resp. $g^{0}_{B}$) if $i=A$ (resp. $i=B$).
\begin{equation}
g^{0}_{\lambda} (E)=\frac{\frac{m_{\lambda}}{m}}{E -g\mu_{\lambda}B/2m - D_{\lambda} \gamma_{\lambda} \frac{m_{\lambda}}{m}}
\end{equation}

where $\lambda = A$ or B. For convenience, we have also introduced the reduced variable $E=\frac{\omega}{2m}$, m denotes the averaged magnetization: $m=\sum_{\lambda} c_{\lambda} m_{\lambda}$. The term which is proportional to $\epsilon$ comes from the environmental disorder term. This term is crucial to recover the Goldstone mode and requires to be treated very carefully. We have introduced the coefficient $\epsilon$ which is in principle equal to 1, in order to follow the influence of the environmental disorder term during  the calculations. Note also that this term appears because of RPA decoupling.
If $\epsilon =0$ Eq. \ref{propag} is analogous to the propagator of an electron in a disordered medium with {\it on-site} potential and random long-range hopping terms $ t_{il}=\Phi_{il}$ ({\it off-diagonal} disorder). In this case the problem can be solved just within the BEB formalism. However, one should stress that the BEB formalism does not apply when the environmental term is present.
Note also that in our model the locator $g^{0}_{\lambda}$, $\Psi_{il}$ and $\gamma_{i}$ are all temperature dependent, thus CPA and RPA loops have to be treated simultaneously in a self-consistent manner. It is also interesting to note that $\Psi_{il} \ne \Psi_{li}$ in the case where the sites i and l are occupied by different type of atoms.

\section{Cumulant expansion method for the averaged Green's functions.}

As it is done in Ref. \cite{Whitelaw}, the basic idea is to write Eq.\ref{propag} as a locator expansion in BEB manner \cite{BEB}. We define the random variable $p_{i}$: $p_{i}=1$ if A is at site i or $p_{i}=0$ if i is occupied by a B ion. Therefore the locator reads,
\begin{equation}
g_i=p_i g_{A}^{0} + (1-p_i) g_{B}^{0}= g^A_i+g^B_i
\end{equation}
and, 
\begin{eqnarray}
\phi_{il}=p_i J^{AA}_{il}p_l +p_i J^{AB}_{il} (1-p_l)+ \nonumber \\
(1-p_i)J^{AB}_{il}p_l+(1-p_i)J^{BB}_{il}(1-p_l)
\end{eqnarray}
similarly,
\begin{eqnarray}
\Psi_{il}=p_i J^{AA}_{il}p_l +p_i J^{AB,1}_{il} (1-p_l)+\nonumber \\
(1-p_i)J^{AB,2}_{il} p_l+(1-p_i)J^{BB}_{il}(1-p_l)
\end{eqnarray}
where $J^{AB,1}_{il} = \frac{m_{B}}{m_{A}}J^{AB}_{il} $ and $J^{AB,2}_{il} = \frac{m_{A}}{m_{B}}J^{AB}_{il}$.

The Green's function are expressed in term of a 2x2 matrix and one gets for the equation of motion,
\begin{eqnarray}
{\bf G}_{ij}=\left(
\begin{array}{cc}
g^{A}_{i} & 0 \\ 0 & g^{B}_{i}
\end{array}
\right)\delta_{ij}+ 
\left(
\begin{array}{cc}
g^{A}_{i} & 0 \\ 0 & g^{B}_{i}
\end{array}
\right)
\sum_{m}\left( \begin{array}{cc}
J^{AA}_{im} & J^{AB}_{im} \\ J^{AB}_{im} & J^{BB}_{im}
\end{array}
\right)
\left(
\begin{array}{cc}
G^{AA}_{mj} & G^{AB}_{mj} \\ G^{BA}_{mj} & G^{BB}_{mj}
\end{array}
\right)\nonumber \\ 
-\epsilon \left(
\begin{array}{cc}
g_{A}^{0} & 0 \\ 0 & g_{B}^{0}
\end{array}
\right)
\left(
\begin{array}{cc}
J^{AB,1} +\sum_{l}(J^{AA}_{il}-J^{AB,1}_{il})p_l & 0  \\ 0 & J^{BB} +\sum_{l}(J^{AB,2}_{il}-J^{BB}_{il})p_l
\end{array}
\right)
\left(
\begin{array}{cc}
G^{AA}_{ij} & G^{AB}_{ij} \\ G^{BA}_{ij} & G^{BB}_{ij}
\end{array}
\right)
\end{eqnarray}

We have defined the variables $J^{AB,1}=\sum_l J^{AB,1}_{il}$ and $J^{BB}=\sum_l J^{BB}_{il}$ .

The aim is to expand this expression into a product of the p factors, which can then be averaged over disorder by expanding into cumulants.
For that purpose we separate out the factors and introduce a new variable $\rho_i$ by $p_i=\rho_i+c$ (where $c_{A}=c$) . The idea is to separate out the virtual crystal part.

\begin{eqnarray}
{\bf g}_i= \rho_i
\left(
\begin{array}{cc}
g_{A}^{0} & 0 \\ 0 & -g_{B}^{0}
\end{array}
\right)+
\left(
\begin{array}{cc}
cg_{A}^{0} & 0 \\ 0 & (1-c)g_{B}^{0}
\end{array}
\right)
\end{eqnarray}

There is still the environmental term which is more difficult to handle.
As it was done by Lage and Stinchcombe \cite{LS}, by converting into {\bf k}-space the calculations become easier to perform.

We define the Fourier transform by,

\begin{equation}
{\bf G}_{\bf kk'}=\sum_{ij} \exp(i{\bf k} \cdot {\bf r}_{i})\exp(-i{\bf k'} \cdot {\bf r}_j){\bf G}_{ij}
\end{equation}

After some manipulation one gets,

\begin{eqnarray}
{\bf G}_{{\bf kk'}}= {\bf G}^{vc}_{\bf k}\rho_{\bf k-k'}+{\bf G}^{vc}_{\bf k}
\left(
\begin{array}{cc}
c & 0 \\ 0 & c-1
\end{array}
\right) \delta_{\bf k-k'}+
{\bf G}^{vc}_{\bf k} \frac{1}{N} \sum_{\bf q}\rho_{\bf k-q}
{\bf V}_{\bf kq}{\bf G}_{\bf qk'}
\label{propag2}
\end{eqnarray}

where the $2 \times 2$ matrix ${\bf V}_{\bf kq}$ is defined by:
\begin{eqnarray}
{\bf V}_{\bf kq}=\left(
\begin{array}{cc}
J^{AA}_{\bf q} -\epsilon (J^{AA}_{\bf k-q}-J^{AB,1}_{\bf k-q}) & J^{AB}_{\bf q} \\ J^{AB}_{\bf q} & J^{BB}_{\bf q} -\epsilon (J^{BB}_{\bf k-q}-J^{AB,2}_{\bf k-q})
\end{array}\right)
\end{eqnarray}

and the virtual-crystal Green's function ${\bf G}^{vc}_{\bf k}$,
\begin{eqnarray}
[{\bf G}^{vc}_{\bf k}]^{-1}={\bf M}_{0}-c{\bf M}_{1}
\end{eqnarray}

where the matrices ${\bf M}_{0}$ and ${\bf M}_{1}$ are,
\begin{eqnarray}
{\bf M}_{0}=
\left(
\begin{array}{cc}
(g_{A}^{0})^{-1} & 0 \\ 0 & -(g_{B}^{0})^{-1}
\end{array}\right)+
\left(
\begin{array}{cc}
\epsilon J^{AB,1} & 0 \\ J^{AB}_{\bf k} & J^{BB}_{\bf k}-\epsilon J^{BB}
\end{array}\right)
\end{eqnarray}
and,

\begin{eqnarray}
{\bf M}_{1}=
\left(
\begin{array}{cc}
J^{AA}_{\bf k}-\epsilon(J^{AA}- J^{AB,1}) & J^{AB}_{\bf k}  \\ J^{AB}_{\bf k} & J^{BB}_{\bf k}-\epsilon (J^{BB}- J^{AB,2})
\end{array}\right)
\end{eqnarray}
The equation (\ref{propag2}) can be expanded into 2 sub-series.
\begin{eqnarray}
{\bf G}_{\bf kk'}={\bf G}^{(1)}_{\bf kk'}+{\bf G}^{(2)}_{\bf kk'}
\end{eqnarray}
where the sub-series are respectively,
\begin{eqnarray}
{\bf G}^{(1)}_{\bf kk'}=
{\bf G}^{vc}_{\bf k}\rho_{\bf k-k'}+
\frac{1}{N}\sum_{\bf q}{\bf G}^{vc}_{\bf k}{\bf V}_{\bf kq}{\bf
G}^{vc}_{\bf q}\rho_{\bf k-q}\rho_{\bf q-k'}+....
\end{eqnarray}
and,
\begin{eqnarray}
{\bf G}^{(2)}_{\bf kk'}=
\left({\bf G}^{vc}_{\bf k}\delta_{\bf k-k'}+{\bf G}^{vc}_{\bf k}{\bf V}_{\bf kk'}{\bf
G}^{vc}_{\bf k'}\rho_{\bf k-k'}+
\frac{1}{N}\sum_{\bf q}{\bf G}^{vc}_{\bf k}{\bf V}_{\bf kq}{\bf
G}^{vc}_{\bf q}{\bf V}_{\bf qk'}{\bf G}^{vc}_{\bf k'}
\rho_{\bf k-q}\rho_{\bf q-k'}+...\right)
\left(
\begin{array}{cc}
c & 0 \\ 0 & c-1
\end{array}\right)
\end{eqnarray}

The averaged Green's function is obtained by averaging over products of $\rho$ by expanding into cumulants $P_{i}(c)$. 
For instance, 
\begin{equation}
\langle \rho_{\bf k_1}\rho_{\bf k_2} \rangle =\frac{P_{2}(c)}{N} 
\delta({\bf k_1}+ {\bf k_2}) 
\end{equation}
\begin{equation}
\langle \rho_{\bf k_1}\rho_{\bf k_2}\rho_{\bf k_3} \rangle =\frac{P_{3}(c)}{N^{2}} 
\delta({\bf k_1}+ {\bf k_2} +{\bf k_3} ) 
\end{equation}
and,
\begin{eqnarray}
\langle \rho_{\bf k_1}\rho_{\bf k_2}\rho_{\bf k_3} \rho_{\bf k_4}\rangle =\frac{P_{4}(c)}{N^{3}} 
\delta({\bf k_1}+ {\bf k_2} +{\bf k_3} ) + (\frac{P_{2}(c)}{N})^{2}[
\delta({\bf k_1} +{\bf k_2})\delta({\bf k_3} +{\bf k_4})+\nonumber \\ 
\delta({\bf k_1} +{\bf k_3})\delta({\bf k_2} +{\bf k_4})+ 
\delta({\bf k_1} +{\bf k_4})\delta({\bf k_2} +{\bf k_3})]
\end{eqnarray}

The cumulants are systematically obtained the generating function,
\begin{equation}
g(x,c)=\text{ln}(1-c+ce^{x})=\sum_{i=1}^{\infty} P_{i}(c) \frac{x^{i}}{i!}
\end{equation}
From this equation one gets $P_{1}(c)=c$, $P_{2}(c)=c(1-c)$, $P_{3}(c)=c(1-c)(1-2c)$....

In order to get a closed form for the series we have to make the usual CPA approximation which consists in keeping only the diagrams with no crossings of external lines. As it is was shown by Yonezawa et al. \cite{Yonezawa,Leath}, the self-consistency requires a modification of the semi-invariants to be attributed to each vertex. In other words it means that the cumulants $P_{i}(c)$ have to be replaced by a new set of coefficients $Q_{i}(c)$ which satisfies the relation,
\begin{equation}
Q_{1}(c)+Q_{2}(c)x+Q_{3}(c)x^2 .....=\sigma_c(x)=\frac{c}{1-x(1-\sigma_c(x))}
\label{sigm}
\end{equation}

where the modified cumulants are,
\begin{equation}
Q_{i}(c)=\sum_{m=1}^{i} [(-1)^{m-1} \frac{(i+m-2)!}{m!(i-m)!(m-1)!}] c^{m}
\end{equation}

In the single site approximation, after averaging, one gets for the averaged $2\times2$ Green's function matrix,

\begin{equation}
{\bf \bar{G}}_{\bf kk'}={\bf \bar{G}}_{\bf k}\delta_{\bf k-k'}={\bf \tilde{G}}_{\bf k}
\left[\left(
\begin{array}{cc}
c & 0 \\ 0 & c-1
\end{array}\right) + {\bf \Delta}_{\bf k}
\right]
\label{fullg}
\end{equation}

where,

\begin{equation}
{\bf \tilde{G}}_{\bf k}=\left[({\bf G}^{vc}_{\bf k})^{-1}-{\bf \Sigma}_{\bf k}
\right]^{-1}
\label{ghat}
\end{equation}
${\bf \Sigma}_{\bf k}$ denotes the self-energy, it is given by,

\begin{equation}
{\bf \Sigma}_{\bf k}=Q_{2} \frac{1}{N} \sum_{\bf q}{\bf V}_{\bf kq}{\bf \tilde{G}}_{\bf q}{\bf V}_{\bf qk}+
Q_{3} \frac{1}{N^2} \sum_{\bf q,t}{\bf V}_{\bf kq}{\bf \tilde{G}}_{\bf q}{\bf V}_{\bf qt}
{\bf \tilde{G}}_{\bf t}{\bf V}_{\bf tk}+...
\label{self}
\end{equation}
and,

\begin{equation}
{\bf \Delta}_{\bf k}=Q_{2} \frac{1}{N} \sum_{\bf q}{\bf V}_{\bf kq}{\bf \tilde{G}}_{\bf q}+
Q_{3} \frac{1}{N^2} \sum_{\bf q,t}{\bf V}_{\bf kq}{\bf \tilde{G}}_{\bf q}{\bf V}_{\bf qt}
{\bf \tilde{G}}_{\bf t}+...
\label{delta}
\end{equation}

The term ${\bf \Delta}_{\bf k}$ which is very similar to the self-energy is called end correction \cite{LS}.
Note that, inside the CPA loop, Eq.(\ref{ghat}) and Eq.(\ref{self}) are the only 2 equations which have to be solved self-consistently.
To summarize, in Fig. \ref{fig1} we show a diagrammatic representation of the previous set of equations.

\subsection{Evaluation of ${\bf \Delta}_{\bf k}$.}

It is convenient for the calculations to start by defining,
\begin{equation}
\gamma_{i}({\bf q})=\frac{1}{z_i} \sum_{{\bf r}^i_l} \exp(i{{\bf qr}^i_l})
\end{equation}

The sum, ${\bf r}^{i}_{l}$ runs over the i-th type of neighbors of the i-th shell 
from a given site 0 and $z_i$ is the total number of neighbors in the shell. Note that, from 
now $\sum_{i}$ will correspond to a summation over the different shells. With this definition it follows immediately,

\begin{equation}
J^{AA}({\bf q})= \sum_{i} J^{AA}_{i} z_{i} \gamma_{i}({\bf q})
\end{equation}
We get similar expression for $J^{BB}({\bf q})$ and $J^{AB}({\bf q})$...

It is convenient to decompose the matrix ${\bf V}_{\bf kq}$ into two terms,

\begin{equation}
{\bf V}_{\bf kq}={\bf V}^{(1)}_{\bf kq}+{\bf V}^{(2)}_{\bf kq}
\end{equation}
where
\begin{equation}
{\bf V}^{(1)}_{\bf kq}=\sum_{i}{\bf V}^{(1),i}_{\bf kq}=
\sum_{i} \left[{\bf A}_{i}-\epsilon {\bf D}_{i}\gamma_{i}({\bf k})\right]\gamma_{i}({\bf q})
\label{vpot}
\end{equation}
and,
\begin{equation}
{\bf V}^{(2)}_{\bf kq}=\sum_{i}{\bf V}^{(2),i}_{\bf kq}=\epsilon {\bf D}_{i}\left[\gamma_{i}({\bf k})\gamma_{i}({\bf q}) - \gamma_{i}({\bf k-q})\right]
\end{equation}

${\bf A}_{i}$ and ${\bf D}_{i}$ are the following 2x2 matrices:

\begin{equation}
{\bf A}_{i}=\left(
\begin{array}{cc}
J^{AA}_i & J^{AB}_i \\ J^{AB}_i & J^{BB}_i
\end{array}\right)z_{i}
\end{equation}

\begin{equation}
{\bf D}_{i}=\left(
\begin{array}{cc}
J^{AA}_i-J^{AB,1}_i & 0 \\ 0 & J^{BB}_i-J^{AB,2}_i
\end{array}\right)z_{i}
\end{equation}
By using the following very useful property \cite{Callen}: if $f(r)$ is a function which is equivaluated at each site $r_i$ of $E_i$ then,
\begin{equation}
\frac{1}{N} \sum_{\bf q}\gamma_{i}({\bf k-q})f({\bf q})=\gamma_{i}({\bf k})\frac{1}{N}\sum_{\bf q} \gamma_{i}({\bf q})f({\bf q})
\label{prop}
\end{equation}

By using Eq.(\ref{prop}), we find significant simplifications in the calculations. Indeed all the terms of the serie involving at least one factor ${{\bf V}}^{(2)}$ reduces to zero. Thus the end correction term does not explicitely depend on the environmental disorder term.

After calculation we finally get,

\begin{equation}
{\bf \Delta}_{\bf k}= \sum_{ij}{{\bf V}}^{(1),i}_{\bf k,0} \left[ Q_{2}{\bf I}+Q_{3} {\bf M}+Q_{4}
{\bf M}^{2}+....
\right]_{ij}{{\bf F}^{j}}
\label{delt}
\end{equation}

Like ${{\bf V}^{(1),i}}$, ${{\bf F}^{j}}$ is a $2 \times 2$ matrix, and ${\bf M}$ a $N_{s} \times N_{s}$ matrix, where each matrix element ${\bf M}_{ij}$ is a $2 \times 2$ matrix. $N_{s}$ denotes the number of considered shells. ${{\bf V}^{(1),i}}$ is given in Eq. (\ref{vpot}) and ${{\bf F}^{j}}$ and ${\bf M}_{ij}$ are defined by
\begin{equation}
{\bf F}^i=\frac{1}{N} \sum_{\bf q}\gamma_{i}({\bf q}){\bf \tilde{G}}_{\bf q}
\end{equation}
and,
\begin{equation}
{\bf M}_{ij}=\frac{1}{N} \sum_{\bf q}\gamma_{i}({\bf q}){\bf \tilde{G}}_{{\bf q}}{\bf V}^{(1),j}_{\bf q,0}
\end{equation}

The sum in Eq. \ref{delt} is obtained after diagonalization of the $2N_{s} \times 2N_{s}$ matrix
${\bf M}={\bf P}^{-1} {\bf M}_{diag} {\bf P}$,
\begin{equation}
Q_{2}{\bf I}+Q_{3}{\bf M}+Q_{4}{\bf M}^{2}+....={\bf P}^{-1} 
[(\sigma_{c}({\bf M}_{diag})-Q_{1}{\bf I}){\bf M}_{diag}^{-1}]{\bf P}
\end{equation}
The function $\sigma_{c}$ was previously defined in Eq. (\ref{sigm}), and $[\sigma_{c}({\bf M}_{diag})]_{ij} = 
\sigma_{c}(\lambda_{i}) \delta_{ij}$ where $\lambda_{i}$ are the eigenvalues of ${\bf M}$.
Hence, we get for the end correction

\begin{equation}
{\bf \Delta}_{\bf k}= \sum_{ij}{{\bf V}}^{(1),i}_{\bf k,0} 
[{\bf P}^{-1} 
[(\sigma_{c}({\bf M}_{diag})-Q_{1}{\bf I}){\bf M}_{diag}^{-1}]{\bf P}]_{ij}
{{\bf F}^{j}}
\end{equation}
Let us now proceed further and evaluate the self-energy ${\bf \Sigma}_{\bf k}$.
\subsection{Evaluation of ${\bf \Sigma}_{\bf k}$.}

Using the remarks made in the previous section, we find that the self-energy can be written,
\begin{equation}
{\bf \Sigma}_{\bf k}={\bf \Sigma}^{(1)}_{\bf k}+{\bf \Sigma}^{(2)}_{\bf k}
\end{equation}

where ${\bf \Sigma}^{(1)}_{\bf k}$ (resp.${\bf \Sigma}^{(2)}_{\bf k}$) is obtained by replacing
${\bf V}_{\bf k,q}$ by ${\bf V}^{(1)}_{\bf k,q}$ (resp.${\bf V}^{(2)}_{\bf k,q}$). Indeed we find that each term of the serie containing both ${\bf V}^{(1)}$ and ${\bf V}^{(2)}$ reduces to zero. After simplifications we obtain for ${\bf \Sigma}^{(1)}_{\bf k}$,
\begin{equation}
{\bf \Sigma }^{(1)}({\bf k})=\sum_{i,j}{{\bf V}^{i}}_{\bf k,0} \left[ Q_{1}{\bf I}+Q_{2}{{\bf M}}+Q_{3}
{\bf M}^{2}+....
\right]_{ij}{\bf \Gamma}^{j}({\bf k})
\end{equation}
where ${\bf \Gamma}^{j}({\bf k})=\gamma_{j}({\bf k}) \left(
\begin{array}{cc}
1 & 0 \\ 0 & 1
\end{array}\right)$. 

As previously done for the end correction, using the function $\sigma_{c}(z)$ defined in Eq. \ref{sigm} we obtain immediately,
\begin{equation}
Q_{1}{\bf I}+Q_{2}{\bf M}+Q_{3}
{\bf M}^{2}+....={\bf P}^{-1} [\sigma({\bf M}_{diag})]{\bf P}
\end{equation}

Note that, we have included in the sum the first order term depending on 
{\it c} ($Q_{1}$) which comes from the virtual crystal Green's function ${\bf G}^{vc}_{q}$.

In general, the evaluation of the second term ${\bf \Sigma }^{(2)}({\bf k})$ is much more complicated. One can get an analytical form only for simple cases. For example if the exchange 
integrals are restricted to only nearest neighbor, the complete summation of the serie can be performed by using the space group symmetry of the lattice \cite{Izyumov,LS}.
In the case of nearest neighbor Heisenberg system one gets,
\begin{equation}
{\bf \Sigma}^{(2)}_{\bf k}(E)={\bf C}_{p}(1-\gamma(2{\bf k}))+{\bf C}_{d}(1+\gamma(2{\bf k})-
2\gamma({\bf k})^{2})
\label{sigm2}
\end{equation}
where,
\begin{equation}
{\bf C}_{p,d}=-\frac{\epsilon}{2}(Q_{1}{\bf I}+Q_{2}{\bf M}_{p,d}+Q_{3}{\bf M}_{p,d}^{2}+...){\bf D}_{1}
\label{cpd}
\end{equation}
${\bf C}_{p,d}$ are evaluated in the same way it was done for ${\bf \Sigma}^{(1)}_{\bf k}(E)$ and
${\bf \Delta}_{\bf k}(E)$. The matrices ${\bf D}_{1}$, ${\bf M}_{p}$ and ${\bf M}_{d}$ are respectively,
\begin{equation}
{\bf D}_{1}=\left(
\begin{array}{cc}
J^{AA}-J^{AB,1} & 0 \\ 0 & J^{BB}-J^{AB,2}
\end{array}\right)z
\end{equation}

\begin{equation}
{\bf M}_{p}=-\frac{\epsilon}{6}{\bf D}_{1}{\bf \tilde{G}}_{p}
\end{equation}

\begin{equation}
{\bf M}_{d}=-\frac{\epsilon}{4}{\bf D}_{1}{\bf \tilde{G}}_{d}
\end{equation}

where ${\bf \tilde{G}}_{p}=\frac{1}{N} \sum_{\bf q} (1-\gamma(2{\bf q})){\bf \tilde{G}}({\bf q})$ and 
${\bf \tilde{G}}_{d}=\frac{1}{N} \sum_{\bf q} (1+\gamma(2{\bf q})-2\gamma({\bf q})^{2}){\bf \tilde{G}}({\bf q})$.

Note that the virtual crystal approximation for ${\bf \Sigma}^{(2)}_{\bf k}(E)$ consists in taking in Eq. (\ref{cpd}) the first term only. Then it follows immediately,
\begin{equation}
C_{p}^{VCA}=C_{d}^{VCA}=-\frac{\epsilon c}{2} {\bf D}_{1}
\end{equation}

which substituted in Eq.(\ref{sigm2}) leads to,
\begin{equation}
{\bf \Sigma}^{(2),VCA}_{\bf k}(E)=-\epsilon c {\bf D}_{1}(1-\gamma({\bf k})^{2})
\label{sig2vca}
\end{equation}
Note that ${\bf \Sigma}^{(2),VCA}_{\bf k}$ is energy independent. It is also important to stress that at the lowest order the self-consistency for ${\bf \Sigma}^{(2)}$ is not required.

Most of the ferromagnetic materials are of itinerant type, which means that the exchange integrals between different localized magnetic ions are long range and driven by the polarization of the conduction electrons gas as it is for the RKKY mechanism \cite{rkky}.
Analytically, the generalization of the previous calculations to the more interesting  case where $J_{ij}$ are long ranged is not an easy task. However by truncating the serie, the summation can be performed numerically. It is important to note that ${\bf \Sigma }^{(2)}({\bf k})$ is (i) proportional 
to $\epsilon$ which means that it originates only from the environmental disorder term and (ii) each term of the serie vanishes in the long wave length limit ${\bf \Sigma }^{(2)}({\bf k=0})=0$. This implies that even after truncation of the serie at any order, the Goldstone theorem remains fulfilled. Thus the long wave length magnons are always treated properly.
Furthermore, since ${\bf \Sigma }^{(2)}({\bf k})$ corresponds to higher order scattering terms (p,d,f,...) it is natural to expect that these terms should not affect the Curie temperature in a dramatic way. In other words we expect that a truncation of ${\bf \Sigma }^{(2)}({\bf k})$ serie to the first few term should already provide a good approximation of Curie temperature compared to the one one would get with the complete series. However, it is crucial to consider at least the lowest order term (the virtual crystal contribution) otherwise even in the clean limit 
one would not  recover the correct result and the Goldstone's theorem would be violated . If we consider the lower approximation 
${\bf \Sigma }^(2) \approx {\bf \Sigma }^{(2)}_{VCA}$, we get the expected results in the limit  $c=0$ and $c=1$. It is not a priori clear whether such an approximation of ${\bf \Sigma }^{(2)}({\bf k})$ to the lowest order provides satisfying results for the Curie temperature at moderate impurity concentration. Such an approximation will be tested later on.

To conclude this section the complete averaged $2 \times 2$ Green's function is obtained
after solving self-consistently the set of equations Eqs. (\ref{ghat}) and (\ref{self}) within the CPA loop and then using Eqs. \ref{fullg} and \ref{delta} to get ${\bf \Delta}_{\bf k}$ and ${\bf {\bar G}_k}$.
However, as was already mentioned in the introduction, the problem is not solved until we are able to calculate the locators $g_{\lambda}^{0}$ and the exchanged integrals $\Psi_{il}$ which depend on the averaged magnetization $m_{\lambda}$. The determination of $m_{\lambda}$ has to be done self-consistently in an additional external loop (RPA loop).

\section{Magnetization and Curie temperature.}

We assume that the averaged $2 \times 2 $ Green's function matrix ${\bf \bar G} ({\bf k},E)$ is calculated according to the previous section within the CPA loop. We show how from ${\bar G}_{\lambda} ({\bf k},E)$, $\lambda =A$ or B we can get the missing self-consistent equations (RPA loop)
to get the temperature dependent locator $g_{\lambda}^{0}$ and the exchange integrals $\Psi_{il}$. This will allow us to calculate the element-resolved magnetizations $m_{\lambda}=<S_{\lambda}^{z}>$ as function of temperature and the Curie temperature. It was shown by Callen, in the case of a clean system (pure A or B) that the magnetization can be expressed in the following way \cite{Callen},
\begin{equation}
m_{\lambda}=\frac{(S_{\lambda}-\Phi_{\lambda})(1+\Phi_{\lambda})^{2S_{\lambda}+1}+(S_{\lambda}+1+\Phi_{\lambda})\Phi_{\lambda}^{2S_{\lambda}+1}}{(1+\Phi_{\lambda})^{2S_{\lambda}+1}-\Phi_{\lambda}^{2S_{\lambda}+1}}
\label{Callen-form}
\end{equation}
where $\Phi_{\lambda}= \frac{1}{N} \sum_{\bf q}\Phi_{\lambda}({\bf q})$ and $\Phi_{\lambda}({\bf q})$ is defined as,
\begin{equation}
\Phi_{\lambda}({\bf q})=\int_{-\infty}^{+\infty} dE \frac{A_{\lambda}({\bf q},E)}{e^{2mE/kT}-1
}
\label{eqfi}
\end{equation}

where,
\begin{equation}
A_{\lambda}({\bf q},E)=\frac{-1}{\pi} \text {Im} G_{\lambda}^{+-}({\bf q},E)
\end{equation}
is the spectral function.

Note also that the Callen's approach to get the magnetization allows to derive a lot of local spin-spin correlation, they are only expressed as a function of $\Phi_{\lambda}$. For instance,
\begin{equation}
\langle (S_{\lambda}^{z})^{2} \rangle = S(S+1) - m_{\lambda}(1+2\Phi_{\lambda})
\label{s2}
\end{equation}
which is needed to determine the anisotropy parameters $\gamma_{\lambda}$ given in Eq. (\ref{aniso}).

In the case of clean systems, the normalized spectral function $A_{\lambda}({\bf q},E)$ is given by
\begin{equation}
A_{\lambda}({\bf q},E)= \delta(E -E({\bf q}))
\end{equation}
$E({\bf q})=\omega({\bf q})/2m$ and $\omega({\bf q})$ denotes the magnon dispersion.

In the case of a binary (or multi-component) alloy this formula can be generalized in the following way,

\begin{equation}
A_{\lambda}({\bf q},E)=\frac{-1}{\pi} \frac{\text{Im} \bar{G}_{\lambda}^{+-}({\bf q},E)}
{c_{\lambda} x_{\lambda}}
\label{spectral}
\end{equation}

where $c_{\lambda}$ is the concentration of $\lambda$-ion and we have for convenience introduced a T-dependent reduced variable $x_{\lambda}=\frac{m_{\lambda}}{m}$.

Note that in the presence of impurities the spectral function is not anymore a $\delta$ function, but because of the finite imaginary part of the self-energy 
will consists of peaks of finite width with more or less a Lorentzian shape. In the case of binary alloy we expect for a given ${\bf q}$, 2 peaks, more generally {\it n} peaks for an {\it n}-component alloy.

For a given temperature the complete self-consistency is obtained by (i) providing good starting 
values for $m_{\lambda}$ then (ii) performing the CPA loop which provide ${\bf \bar G} ({\bf k},E)$ and finally (ii) going into the RPA loop by using Eq. (\ref{Callen-form}), (\ref{s2}) and (\ref{spectral}) one gets the new values of $m_{\lambda}$ and $\langle (S_{\lambda}^{z})^{2} \rangle$ which are re-injected in the locators $g_{\lambda}^{0}$, the exchange integrals $\Psi_{il}$ and $\gamma_{\lambda}$.

Let us now show how to get the Curie temperature of a disordered Heisenberg binary alloy.
We start by expanding Eq. (\ref{eqfi}) in the limit $T \rightarrow T_{c}$  (i.e, $m_{\lambda} \rightarrow 0$). We immediately get,

\begin{equation}
\Phi_{\lambda} \approx  \frac{kT_{c}}{2m} F_{\lambda}
\end{equation}
where,
\begin{equation}
F_{\lambda}=\frac{1}{N}\sum_{\bf q}
\int_{-\infty}^{+\infty} dE \frac{A_{\lambda}({\bf q},E)}{E}
\end{equation}

After expanding Eq. \ref{Callen-form} as a function of $\frac{1}{\Phi_{\lambda}}$ one obtains,

\begin{equation}
m_{\lambda}=\frac{S_{\lambda}(S_{\lambda}+1)}{3}\frac{2m}{kT_{c}}\frac{1}{F_{\lambda}}
\end{equation}

Since the averaged magnetization m is defined by, $m=\sum_{\lambda}c_{\lambda}m_{\lambda}$, combining the two previous equations one finds for the Curie-Temperature,

\begin{equation}
k_{B}T_{c}=\frac{2}{3}\sum_{\lambda}c_{\lambda} \frac{S_{\lambda}(S_{\lambda}+1)}{F_{\lambda}}
\label{eqtc}
\end{equation}

 Eq. (\ref{eqtc}) is the RPA generalization of the Curie Temperature to a multi-component disordered alloy. The previous equation provides a direct measure of the weight $w_{\lambda}=\frac{1}{k_{B}T_{c}}[c_{\lambda}\frac{S_{\lambda}(S_{\lambda}+1)}{F_{\lambda}}]$ of each $\lambda$-element to the Curie Temperature.

\section{Numerical results.}

In this section we provide an illustration of the RPA-CPA theory and a test for the approximation suggested above for the higher order scattering contribution of the self-energy.
For simplicity, we consider the case of a 3D disordered binary alloy on a simple cubic lattice.
Additionally we restrict the exchange integrals to  nearest neighbor only which allows us to test the validity of the approximation scheme suggested in Sec III Bfor estmating $\Sigma^{2}$.
For further simplifications of the calculations we consider the cas of a zero external field and neglect the anisotropy term which is reasonable for a 3D systems.

In Fig. \ref{fig2}, we have plotted the Curie Temperature as a function of $c$ obtained  with the full CPA treatment, the $\Sigma^{(2)}$ part of the self-energy is calculated exactly (full summation of the serie). Note that pure A (resp. B) corresponds to $c=1$ (resp. $c=0$). Depending on the chosen set of parameters $T_{c}$ shows (i) a minimum ($J_{AB} S_{A} S_{B} \le \text{min}(J_{AA} S_{A}^{2},J_{BB} S_{B}^{2})$, (ii) a maximum ($J_{AB} S_{A} S_{B} \ge \text{max}(J_{AA} S_{A}^{2},J_{BB} S_{B}^{2})$ or (iii) is monotonic ($ \text{min}(J_{AA} S_{A}^{2},J_{BB} S_{B}^{2}) \le J_{AB} S_{A} S_{B} \le \text{max}(J_{AA} S_{A}^{2},J_{BB} S_{B}^{2})$. These three different cases are shown in the figure.

As already mentioned in section III, it is difficult to perform the full summation of
$\Sigma^{(2)}$ for the case of long-range exchange integrals which is the case of many realistic and interesting systems, for example permalloy. As it was discussed previously the simplest approximation consists in keeping only the lowest order term of the serie ( virtual crystal approximation).
In the case of nearest neighbor Heisenberg system $\Sigma^{(2)}$ and  $\Sigma^{(2),VCA}$ are respectively given in Eqs (\ref{sigm2}) and (\ref{sig2vca}).
In Fig. \ref{fig3} we have plotted the Curie Temperature calculated with (i) a full CPA treatment, (ii) the one performed with the approximation discussed previously ${\bf \Sigma}^{(2)}={\bf \Sigma}^{(2)}_{VCA}$ and (iii) the one obtained with virtual crystal approximation.
In the case (iii), the averaged Green's function is,
\begin{equation}
{\bf \bar{G}}_{\bf k}={\bf G}^{vc}_{\bf k}
\left(
\begin{array}{cc}
c & 0 \\ 0 & c-1
\end{array}\right)
\end{equation}
since in  VCA  ${\bf \Delta_{k}}=0$.

The comparison between the full CPA and the virtual crystal approximation shows that the Curie temperature differs significantly. Even, very close to the clean limit the VCA appears to be inappropriate, for instance for $c=0.1$, we observe that $T^{VCA}_{c}$ is about $35 \%$ larger than the full CPA calculated one. Note that the disagreement is even more pronounced in the vicinity of $c=0$ than $c=1$. This can be understood in the following way: since $J_{AB}=3 J_{BB}=1.5 J_{AA}$ and $S_{A}=S_{B}$ a substitutition of a B site by a A site (close to $c=0$) introduces a change of energy (with respect to the pure case) 2 times larger than a substitution of a B site by a A site near $c=1$.
As discussed previously it is interesting to compare the Curie temperature where the VCA is only done on ${\bf \Sigma}^{(2)}$ ($T^{2,VCA}_{c}$). We observe a good agreement between the full CPA calculated $T_{c}$ and $T^{2,VCA}_{c}$, in the whole range of concentration, the agreement is even excellent for $c \ge 0.6$. A comparison between $T^{VCA}_{c}$ and $T^{2,VCA}_{c}$ in the vicinity of $c=0$ and $c=1$ shows that the reason why the VCA approximation breaks down is essentially because of the crude approximation of the s-part of the scattering. Thus this figure validates a simple treatment of ${\bf \Sigma}^{(2)}$. It is also expected that including only few additional terms of the serie will lead to an excellent agreement in the whole range of concentration. Additionally, the approximation ${\bf \Sigma}^{(2)} \approx {\bf \Sigma}^{(2),VCA}$ will get better in the case of long range exchange integrals.

In Fig. \ref{fig4} we show the temperature dependence of the element-resolved magnetizations. In order to demonstrate the versatility of our approach, we have chosen a set of parameters which mimics a ferrimagnetic behavior with compensation point.
Additionally, the parameters are such that $T_{c}^{A} \gg T_{c}^{B}$.
While the temperature dependence of $m_{A}$ follows a standard behavior, $m_{B} (T)$ start to strongly decrease even at low temperature. For example at $T \approx 2.5$,
$m_{A}$ has reduced by less that $20 \%$ while $m_{B}=0.5 \,m_{B}(0)$.
As a result of our choice of the parameters we see that the averaged magnetization $m_{av}=|c_{A} m_{A}+c_{B} m_{B}|$ is non monotonic and vanishes for an intermediate temperature value (compensation point). It is found that the function $\frac{m_{B}}{m_{A}}(T)$ decreases monotonically with temperature. As a result and since at $T=0$, $\frac{m{B}}{m_{A}}=\frac{S_{B}}{S_{A}}$, thus if $\frac{S_{B}}{S_{A}} \le \frac{c_{A}}{c_{B}}$ then $m_{av}$ will not have a compensation point . However, the condition that $\frac{S_{B}}{S_{A}} \ge \frac{c_{A}}{c_{B}}$ is not sufficient to get one, it also required that $\frac{m_{B}}{m_{A}}(T_{c}) \le \frac{c_{A}}{c_{B}}$.

In Fig. \ref{fig5} we now show the magnon spectral density (MSD) $\rho_{\lambda} (E)=\frac{\text{Im} G^{\lambda} (E)}{x_{\lambda} c_{\lambda}}$ as a function of E. We consider 3 different cases: almost clean A and B ( (a) and (c)), and the intermediate situation $c_{A}=c_{B}=0.5$.
In both, Fig. \ref{fig5}a and Fig. \ref{fig5}c we observe that the MSD is very similar to the clean case. This is clearer in case (c) than (a), it is easy to understand that when doping A with B the difference in energy with the undoped case is only of order $10 \%$ ($J_{AA} (S_{A})^2= 0.8$ and $J_{AB}S_{A}S_{B}= 0.9$) whilst doping B with A the change is more drastic (about $100 \%$).
To get a similar MSD to Fig.\ref{fig5}c for a weakly doped B sample, one should take $c \approx 0.005$.

In Fig. \ref{fig6} we show the spectral function $S_{\lambda} ({\bf q},E)$ as afunction of energy for different values of the momentum ${\bf q}$. This quantity is more interesting that the integrated MSD since it provides direct information about the elementary excitation dispersions and their spectral weight. Additionally it is directly related to inelastic neutron scattering measurements. Let us now briefly discuss Fig. \ref{fig6}. At precisely ${\bf q}={\bf 0}$ momentum, in both $S_{\lambda=A,B}$, we observe 2 peaks structure (i) a well defined peak \cite{footnote1} at $E=0$ as expected since our theory fulfills the Goldstone theorem and (ii) a very broad one at intermediate energy $E \approx 1$.
For intermediate values of the momentum, it is difficult to separate the peak and one get a single broad peak. We see clearly that the peaks are crossing each other at ${\bf q} \approx \frac{\pi}{2} (1,1,1)$. Note that due to (i) the different spectral weight of the peaks and
to the closeness of their location, the single peak-structure which is observed at $ q =\frac{\pi}{2}$ is located at different energy for A and B. From this figure we see also that the dispersion of the second peak is almost flat $E_{2} (q) \approx 1$, while the Goldstone mode $E_{1}({\bf q})$ \cite{footnote2} goes from $E=0$ to $E_{max} \approx 2$ when moving in the $(1,1,1)$-direction. 

\section{Conclusion.}

In conclusion, we have presented in this paper a theory based on Green's function formalism to study magnetism in disordered Heisenberg systems with long range exchange integrals. The disordered Green's function are decoupled within Tyablicov procedure and the disorder ({\it diagonal, off-diagonal} and {\it environmental}) is treated with a $2 \times 2$ modified cumulant CPA approach. 
The crucial point is that we are able to treat simultaneously and self-consistently the RPA and CPA loops. Our theory allows in particular to calculate Curie temperature, spectral functions and temperature dependence of the magnetization for each element as a function of concentration of impurity. Additionally, we have proposed a simplified treatment of the p,d,f .. contribution of the self-energy which is difficult to handle in the case of long range exchange integrals. The approximation was tested successfully on 3D disordered nearest-neighbor Heisenberg systems.
Combined with first principle calculations which can provide the exchange integrals this method appears to be very promising to study magnetism in disordered systems.

%
%
\begin{figure}
\caption[]{
Diagrammatic representation of the averaged Green's function calculated within the CPA loop.
${\bar{\bf G}}$ is the total averaged Green's function, ${\bf \Sigma_{k}}$ is the self-energy and ${\bf \Delta_{k}}$ the end-correction.
}
\label{fig1}
\end{figure}

\begin{figure}
\caption[]{
Curie temperature $T_{c}$ for disordered nearest neighbor Heisenberg ferromagnet as a function of the impurity concentration c (A). The parameters are $S_{A}=2$, $S_{B}=3$, $J_{AA}=-0.2$ and $J_{BB}=-0.15$. We have chosen 3 different values for $J_{AB}$.
}
\label{fig2}
\end{figure}

\begin{figure}
\caption[]{Comparison between the Curie temperature calculated as function of the impurity concentration for a nearest neighbor Heisenberg Ferromagnet.
(a) the full CPA calculation, (b) approximation for ${\bf \Sigma}^{(2)}={\bf \Sigma}^{(2)}_{VCA}$ and (c) the virtual crystal calculation. The chosen set of parameters are written in the figure.
}
\label{fig3} 
\end{figure}

\begin{figure}
\caption[]{
Magnetizations $m_{A}$, $-m_{B}$ and averaged one $c_{av}=|c_{A} m_{A} +c_{b} m_{B}|$ as a function of temperature. The spins are $S_{A}=1$ and $S_{b}=3$, the exchange couplings are $J_{AA}=-1.2$, $J_{BB}=-0.10$ and an anti-ferromagnetic coupling between A and B is taken $J_{AB}=0.15$.
The concentration of A-atoms is $c_{A}=0.70$. 
}
\label{fig4}
\end{figure} 

\begin{figure}
\caption[]{
Density of state $\rho_{\lambda} (E)=\frac{\text{Im} G^{\lambda} (E)}{x_{\lambda} c_{\lambda}}$ as a function of E. The continuous line corresponds to $\lambda = A$ and the dashed line to $\lambda = B$, for 3 different concentration of A: $c=0.05$, 0.5 and 0.95.
The parameters are $S_{A}=2$, $S_{b}=3$, $J_{AA}=-0.2$, $J_{BB}=-0.05$, $J_{AB}=-0.15$ and $T \approx T_{c}$.
}
\label{fig5}
\end{figure}

\begin{figure}
\caption[]{Spectral function $ S_{\lambda}({\bf q},E)=-\frac{1}{\pi} \text{Im} G^{\lambda} ({\bf q},E)$ as a function of $E$ for different momentum ${\bf q}$ where ${\bf q} = q(1,1,1)$. The continuous line corresponds to $\lambda=A$ and the dashed line to $\lambda=B$.
The spins are $S_{A}=2$ and $S_{b}=3$, the exchange couplings are $J_{AA}=-0.2$, $J_{BB}=-0.10$, $J_{AB}=-0.15$ and $c_{A}=0.50$. We have taken $T \approx T_{c}$. For clarity of the picture a small imaginary part $\eta=0.1$ have been added.}
\label{fig6}
\end{figure}


\begin{references}
\bibitem{footnote1} For convenience a small broadening has been introduced to make the figure easier for the reader. At exactly $q=0$ we find that the peak at $E=0$ is a $\delta$-peak consistent with the fact that 
$\text{Im} G_{\lambda} (E=0)=0$.
\end{references}
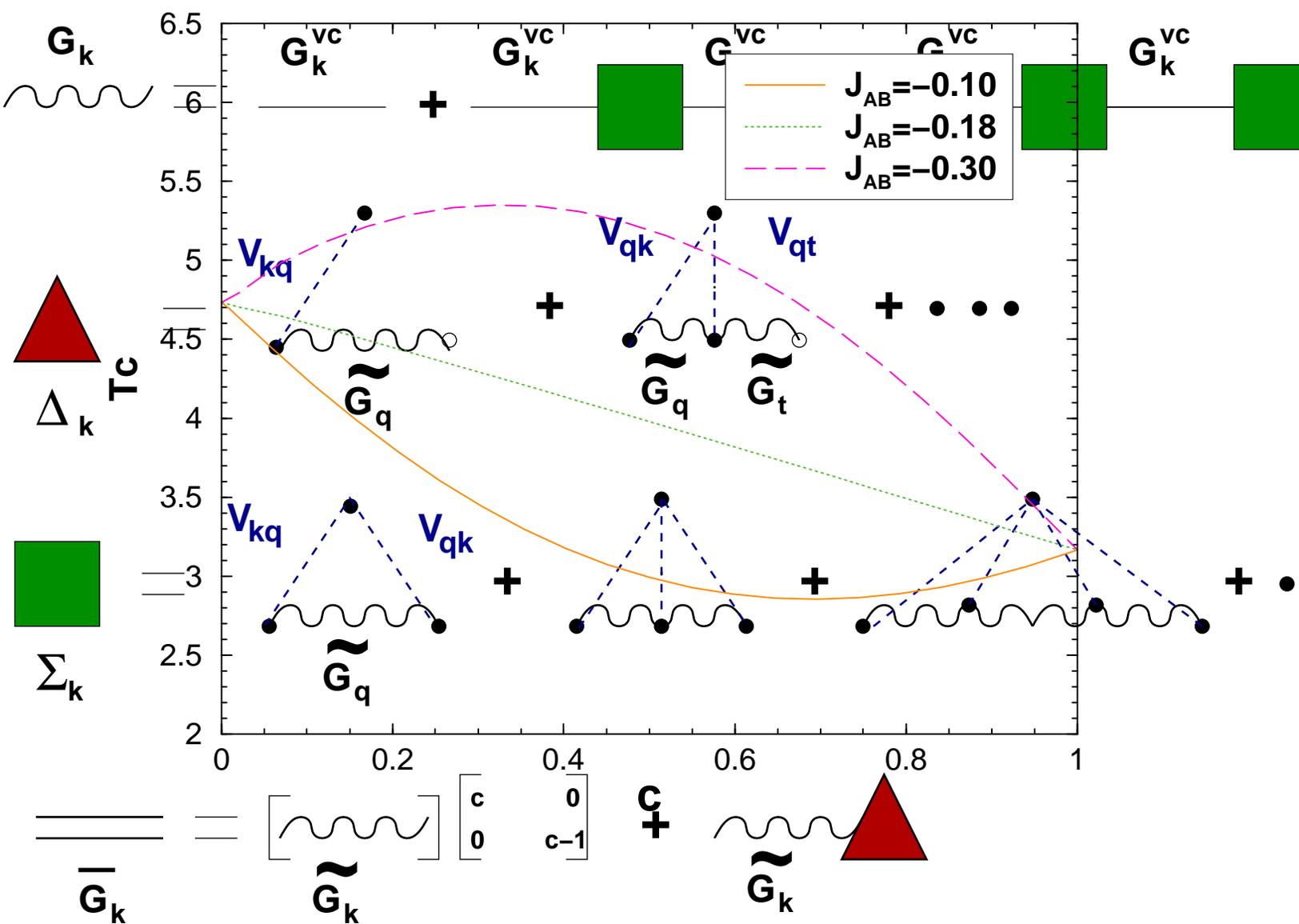
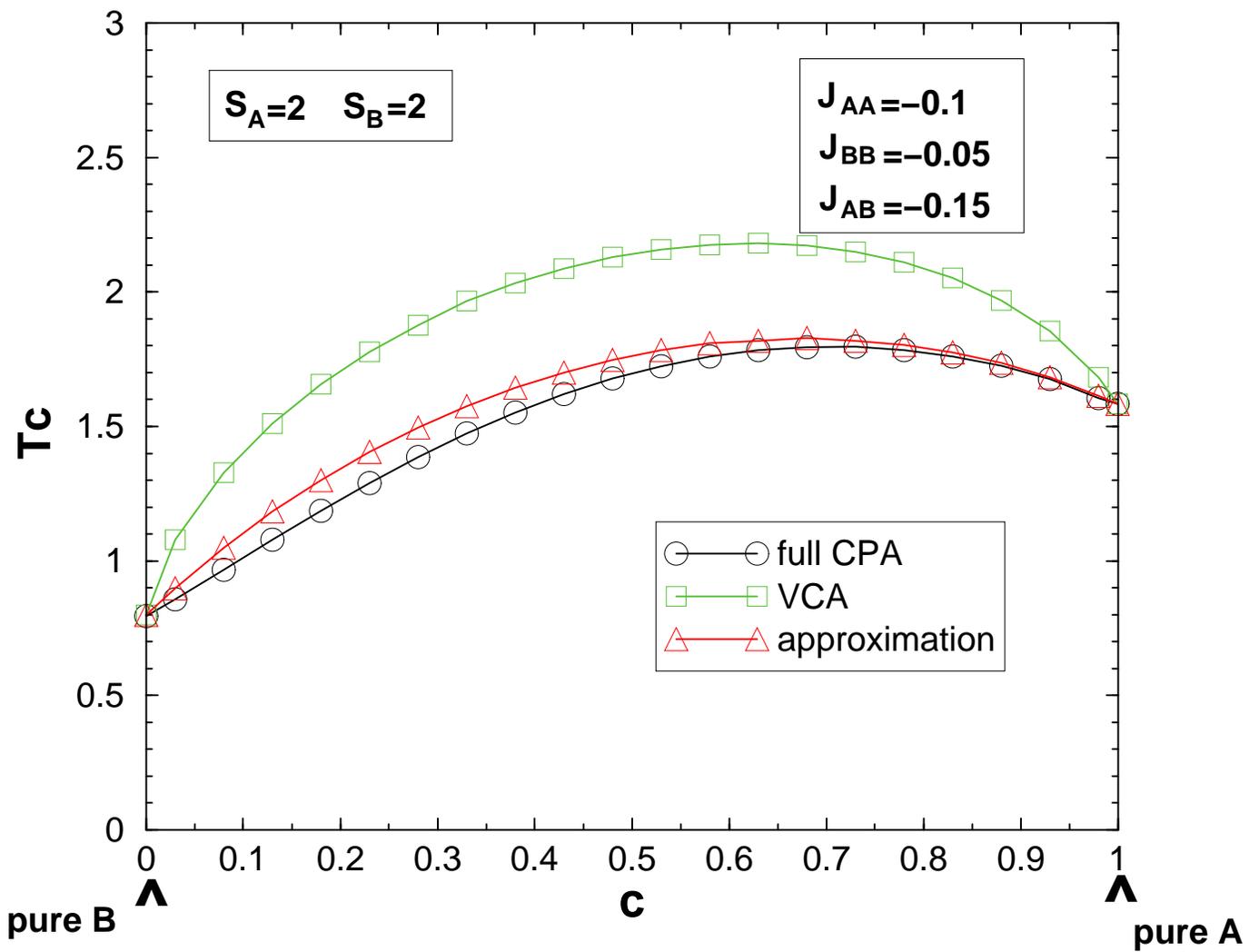
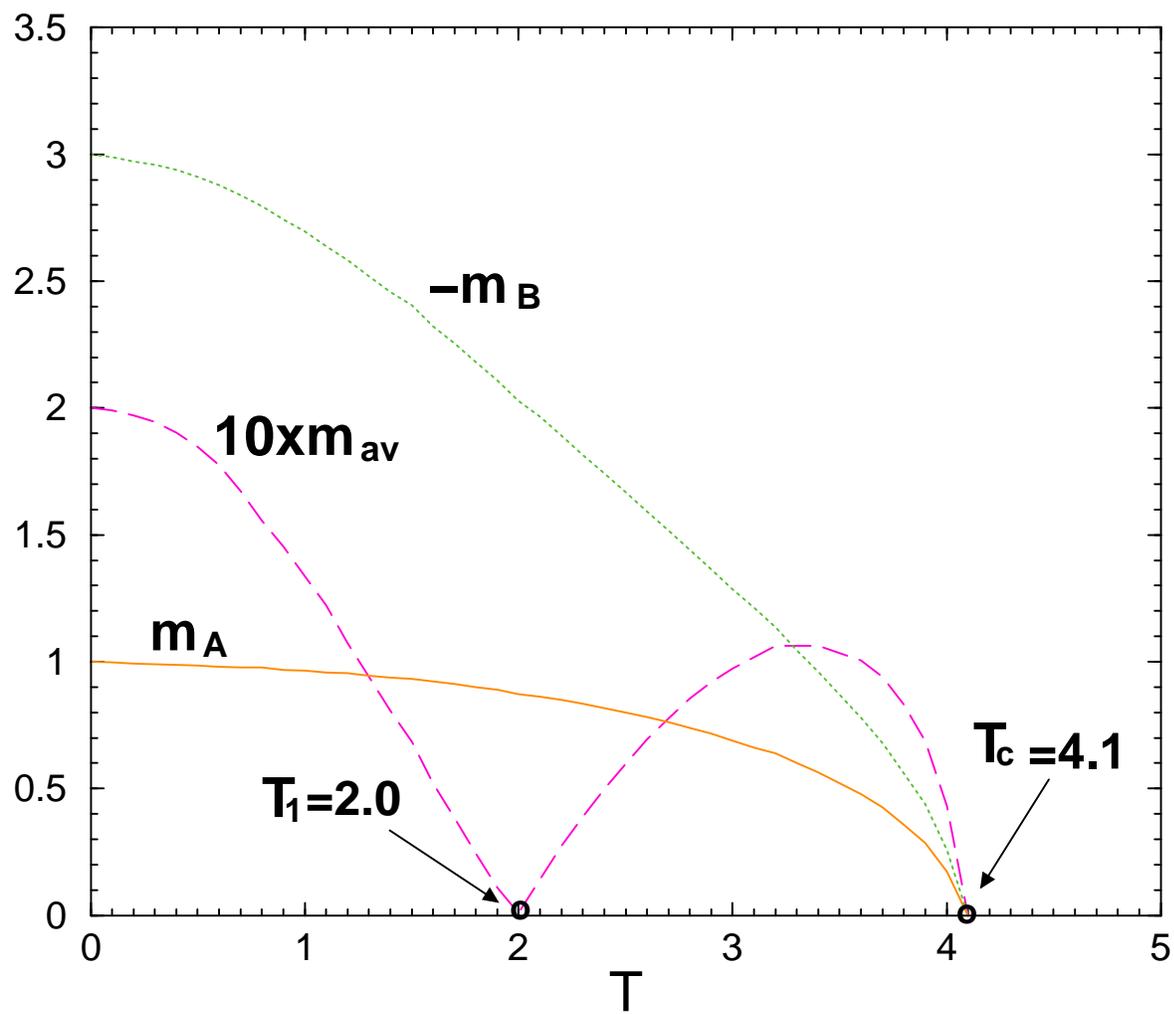
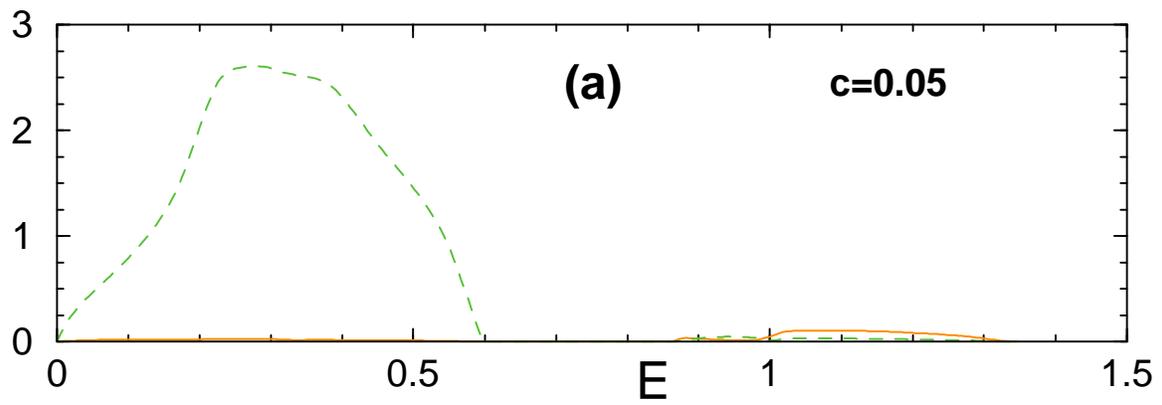
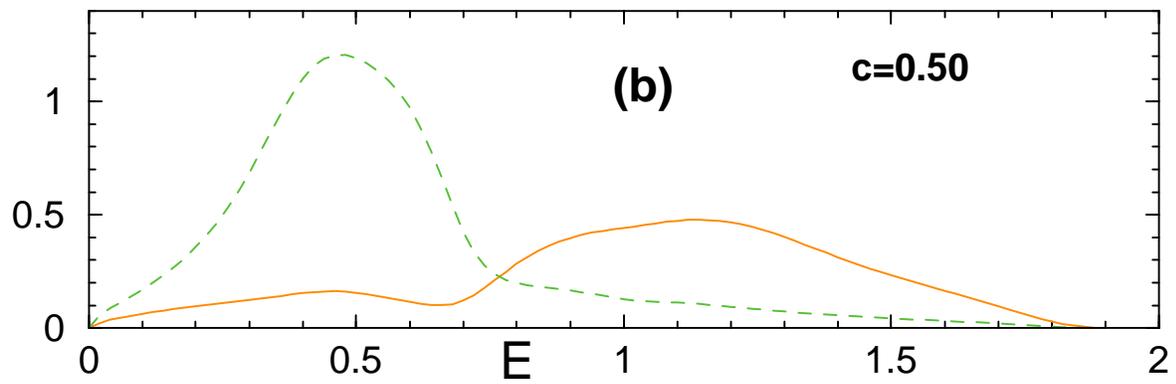
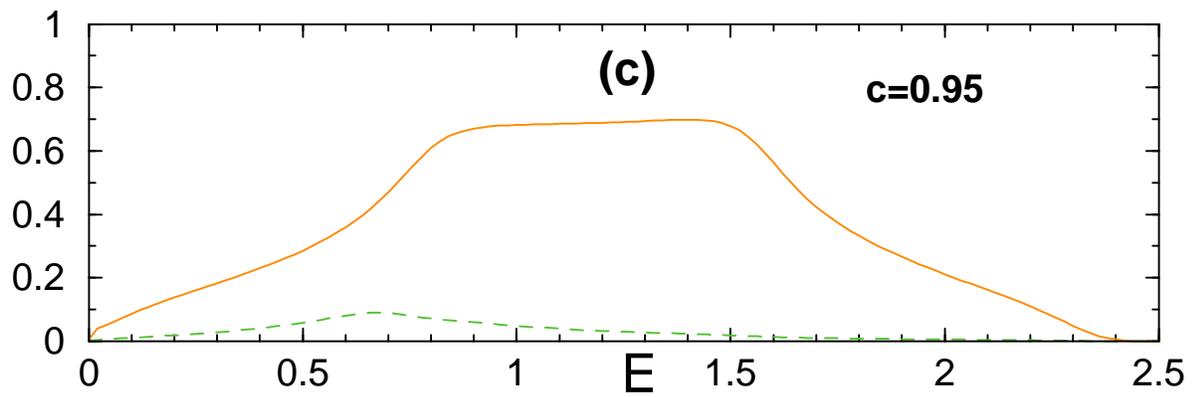
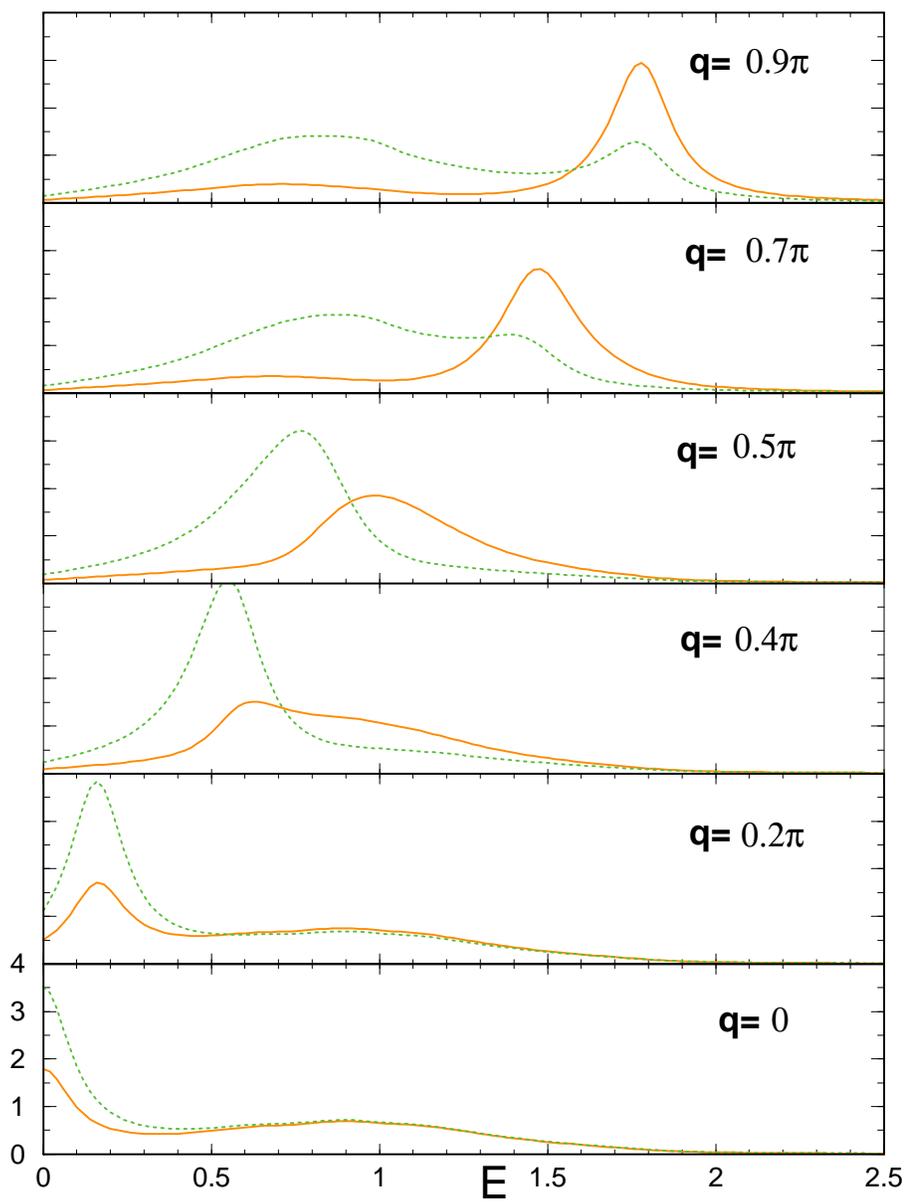
\end{document}